\RequirePackage{etoolbox}
\csdef{input@path}{%
 {sty/}
 {img/}
}%
\csgdef{bibdir}{bib/}
\documentclass[ba]{imsart}

\pubyear{2016}
\volume{00}
\issue{0}
\doi{0000}
\firstpage{1}
\lastpage{1}

\usepackage{color,amsmath}
\usepackage{natbib} 

\begin{document}
\title{Comments on ``Bayesian Solution Uncertainty Quantification for Differential Equations" by Chkrebtii, Campbell, Calderhead \& Girolami}
\author{Jon Cockayne}

\newcommand{\todo}[1]{\textcolor{red}{[TODO: #1]}}

\maketitle

I would like to thank the authors for their interesting and very clearly presented paper discussing probabilistic solvers for ODEs and PDEs. 

\section{Nature of the Uncertainty Quantification}

I am particularly interested in the nature of the uncertainty quantification provided over the forward model. Considering the ODE
\begin{equation*}
	\frac{\textrm{d} u}{\textrm{d} t}(t) = f(t, u) \;,
\end{equation*}
we note that \cite{Skilling:1992fb} advocates construction of a probabilistic model for the vector field $f(t, u)$, the uncertainty of which is then propagated to the solution $u$ itself. This is ``Bayesian'' in that all evaluations of $f$ are incorporated into the estimate of $u$.

Conversely in this work it seems that there is an inconsistency in the posterior distributions obtained. To consider a simply toy example, suppose we wish to solve the linear ODE
\begin{equation*}
	\frac{\textrm{d} u}{\textrm{d} t}(t) = f(t)
\end{equation*} 
where $f$ is independent of $u$, and the problem is thus linear. For a Bayesian treatment of this problem, we endow $u$ with a prior and update it based on evaluations of the vector field $f(t)$ at different $t_i$, $i=1,\dots,N$, where $t_i > t_{i-1}$. If we suppose $u_1 \stackrel{d}{=} u | (t_1, f(t_1))$ and $u_2 \stackrel{d}{=} u | (t_1, f(t_1)), (t_2, f(t_2))$ then we do not expect $u_1(t_1)$ is equal in distribution to $u_2(t_1)$, as a result of having obtained more information about the vector field in $u_2$ which would have an impact on our belief about the distribution $u_2(t_1)$.

However in the present work, it is impossible for $u_1$ to depend upon $f(t_2)$; that is, our new beliefs about $f$ at $t_n$ cannot have any impact on the distribution of $u_m$ for $m < n$. Thus we appear to have imposed a filtration on the $\sigma$-algebra of the probability space which is not inherent to the problem. As a result the posterior distributions cannot be regarded as a full Bayesian update, which I believe this casts some doubt on the ``Bayesian'' nature of the UQ provided in the \cite{Skilling:1992fb} sense, as well as on the information efficiency of the method. 
 
The work is similar to the recently published work of \cite{KerHen16} and \cite{Schober2016}, in that the uncertainty is generated by a methodology similar to ``filtering'' in the data assimilation literature; the full Bayesian posterior would be given by solution of the correspond ``smoothing'' problem.
I would be interested to see whether this can be incorporated into the present work.
 
\section{Treatment of Partial Differential Equations} 
 
The treatment of evolutionary PDEs is also of interest, in light of recent developments of probabilistic meshless methods (PMM) for PDEs \citep{Cockayne2016}. In Sec.~5.3 I was interested to see the reduction of the Navier Stokes PDE to a large system of ODEs.  It is an interesting point for probabilistic numerics, that many problems can be formulated by multiple equivalent numerical schemes; one wonders how the solution obtained by solving this system of ODEs would compare to direct solution of a PDE system, and how consistent the posterior measures generated would be.  

Similarly, in Sec.~5.4 the authors have solved the heat equation by a ``forward in time, continuous in space'' formulation; if I understand this correctly, we treat the spatial component by a Gaussian process model and discretise the temporal component using the methods of this paper. In light of my comments on the provided uncertainty quantification and considering that this evolutionary system is linear, I wonder how this solution would compare to the fully Bayesian solution provided by the PMM.

\bibliographystyle{ba}
\bibliography{refs}
\end{document}